%% file: iclr2027_conference.tex
\documentclass{article} 
\usepackage{iclr2027_conference,times}

\input{math_commands.tex}

\usepackage{hyperref}
\usepackage{url}
\makeatletter
\g@addto@macro\UrlBreaks{\do\0\do\1\do\2\do\3\do\4\do\5\do\6\do\7\do\8\do\9}
\makeatother
\usepackage{booktabs}     
\usepackage{amsmath,amssymb}
\usepackage{amsthm}

\usepackage{graphicx}
\usepackage{xspace}

\usepackage{xcolor}
\usepackage{colortbl}
\usepackage{multirow}
\usepackage{diagbox}
\usepackage{float}
\usepackage{stfloats}
\fnbelowfloat 
\usepackage{wrapfig}
\usepackage{algorithm}
\usepackage{algpseudocode}
\usepackage{enumitem}
\usepackage{soul}
\setul{2pt}{0.5pt}

\setlist{nosep, leftmargin=*} 
\usepackage{tcolorbox} 
\usepackage[disable, colorinlistoftodos]{todonotes}
\setuptodonotes{tickmarkheight=0.2cm}

\makeatletter
\renewcommand{\fps@figure}{tbp}        
\renewcommand{\fps@table}{tbp}         
\makeatother

\title{BoundaryMORPH: Budgeted Reranking via Active Set Selection for Diffuse Retrieval}

\author{%
\makebox[\dimexpr\textwidth-2\tabcolsep\relax][c]{\begin{tabular}[t]{c}
Eylon Caplan\thanks{Research done during internship at AWS AI Labs}~~$^\spadesuit$ \enspace Shamik Roy$^\clubsuit$ \enspace Shib Sankar Dasgupta$^\clubsuit$\\
Yingfan Wang$^\clubsuit$ \enspace Rashmi Gangadharaiah$^\clubsuit$\\
\mdseries $^\spadesuit$Purdue University \qquad $^\clubsuit$AWS AI Labs\\
\mdseries \texttt{ecaplan@purdue.edu}\\
\mdseries \texttt{\{royshami,shibdg,yingfanw,rgangad\}@amazon.com}
\end{tabular}}}

\newcommand{\fce}{f_{\mathrm{CE}}}

\definecolor{agentteal}{rgb}{0.0,0.42,0.45}
\newcommand{\ct}[1]{\textcolor{agentteal}{#1}}

\newif\ifarxiv
\arxivtrue
\ifarxiv\iclrfinalcopy\fi
\begin{document}

\maketitle
\ifarxiv\lhead{Preprint. Under review.}\fi

\begin{abstract}
Open-ended queries in modern Retrieval-Augmented Generation (RAG) are increasingly ``diffuse,'' requiring a large set of documents to be assembled into a finite LLM context window. 
To ensure retrieval quality, systems use fast dual-encoders and more expensive cross-encoders (CEs) to score candidates.
However, the CE budget $B$ is strictly bounded by latency and is often smaller than the context window capacity $k$. 
This mismatch makes standard reranking structurally flawed: it wastes compute verifying obvious top candidates while ignoring relevant documents further down the initial ranking. 
To address this, we introduce BoundaryMORPH, a novel algorithm that allocates CE budget specifically for the LLM's context capacity $k$. 
Using a Gaussian Process, BoundaryMORPH treats the initial dual-encoder ranking as a structural prior and intelligently spends CE calls on resolving top-$k$ set membership at the boundary, rather than seeking a single most-relevant document. 
Information from each CE call propagates to unscored documents, maximizing the utility of the budget. 
We demonstrate that BoundaryMORPH achieves state-of-the-art set retrieval quality across multiple models and datasets with open-ended queries ($+5.4$ nCG@100 over the strongest baseline).
\end{abstract}

\section{Introduction}
\label{sec:intro}

\begin{figure}[b]
    \centering
    \includegraphics[width=\textwidth]{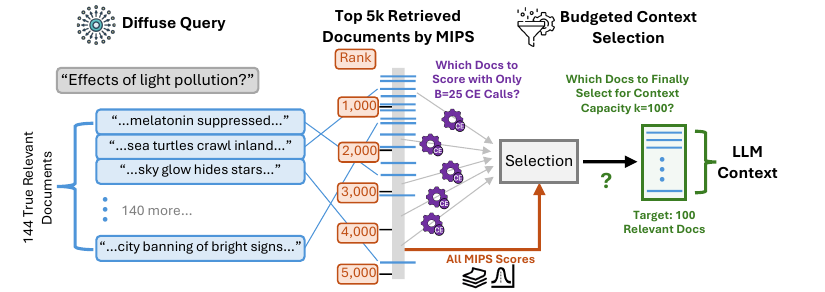}
    \caption{The diffuse query has \emph{many} gold relevant documents (144), and
    they are spread across the MIPS top-5k ranking. The LLM
    generator has been allotted a context capacity $k = 100$, but we have only a CE budget of
    $B = 25$. The selection problem then becomes: how should we optimally spend
    these CE calls and finally select the 100 most relevant documents for the LLM?}
    \label{fig:problem}
\end{figure}

The challenge of serving queries whose relevant material is scattered across many documents has long been recognized in IR \citep{bates_design_1989, marchionini_exploratory_2006,
clarke_novelty_2008}, and they have grown more prominent as people bring broader,
more open-ended requests to LLM-based systems \citep{suri_use_2024,
spatharioti_effects_2025} and on to deep-research agents that survey and
synthesize many sources per query \citep{huang_deep_2025,
coelho_deepresearchgym_2025}. For RAG, this takes the form of queries with
\emph{many} relevant documents; for example, in a deep research setting, a query
like ``Effects of light pollution?'' is expected to surface tens, if not hundreds
of relevant documents (see Figure~\ref{fig:problem})! Such \textbf{diffuse}
queries cannot be well-responded to by referencing a single, or even handful of
relevant documents; rather, they require a larger set of relevant documents to be
inserted into the LLM generator's context for the best response.

Modern RAG systems typically rest on a two-stage architecture: a cheap
\textbf{dual-encoder} embeds the query and every document into a shared vector space and
returns a candidate pool ranked via \textbf{Maximum Inner Product Search} (MIPS) \citep{hambarde_information_2023, zhao_dense_2024, huang_survey_2026}, and an expensive reranker, such as a \textbf{cross-encoder} (CE) or LLM, re-scores a small number of top candidates far more accurately \citep{gao_retrieval-augmented_2024, fan_survey_2024}. 

Because the CE is
an expensive test-time process, it is typically applied to only a limited number,
$B$ (budget calls), of candidates. Likewise, a RAG system is always limited by its
\emph{context capacity}, $k$: the size of the context that has been allotted for
documents being inserted into the RAG generator's context before generating a
response \citep{cuconasu_power_2024, shi_replug_2024, liu_lost_2024,
louis_pisco_2025}.

These two limits become problematic for diffuse queries, for two reasons: (1)
re-scoring only the top $B$ dual-encoder documents risks missing out on relevant
documents further down, and (2) when the CE budget is smaller than the context
capacity ($B < k$), the remaining $k-B$ documents cannot anyway be evaluated by the CE
due to budget constraint, and thus we gain no information about which documents to fit 
into the context, and which to leave out.

Figure~\ref{fig:problem} shows this clearly, where we assume a context capacity of
$k = 100$, but only a $B = 25$ CE call budget: the query is diffuse, with $144$
truly relevant documents, spread across the MIPS ranking. The
context capacity $k = 100$ means we have $100$ slots to fill our context with
(hopefully) $100/144$ relevant documents. However, our budget allows for only $25$
CE calls, which demonstrates problem (2). This begets a resource allocation
problem: how should we optimally \emph{spend} these CE calls and finally
\emph{select} the $100$ most relevant documents for the RAG generator?

To address this, we introduce \textbf{BoundaryMORPH}, a Gaussian Process (GP)
selector \citep{williams_gaussian_1995, srinivas_information-theoretic_2012} that actively estimates the CE's relevance landscape to optimally
allocate the budget of $B$ calls. The use of a GP allows us to \emph{propagate}
information from each CE call, so that even with only, say, $25$ CE calls, we may
still gain information about far more than $25$ documents. BoundaryMORPH is built
on two core insights regarding the nature of diffuse queries and the RAG pipeline:

\begin{enumerate}
\item \textbf{Morphing the MIPS's ranking rather than discarding it:}
  While the cheap dual-encoder is imperfect and locally miscalibrated, it provides
  a strong, scale-free global surface. This signal should be leveraged as a
  structural baseline rather than discarded.
\item \textbf{Targeting the capacity boundary rather than the maximum:} Because
  the RAG context capacity $k$ is known in advance, spending budget to confidently
  confirm the absolute highest-scoring documents is wasteful; they will be
  included in the context regardless. Genuine uncertainty only exists at the
  \textbf{boundary} of inclusion.
\end{enumerate}

BoundaryMORPH operationalizes these insights through a single running GP
estimator. To satisfy the first insight, we set the GP prior mean to the
document's MIPS percentile rank. This allows the
model to smoothly fall back to the dual encoder's judgment in unscored regions,
spending calls only to \emph{bend} this surface rather than rebuild it. To satisfy
the second, BoundaryMORPH abandons traditional global-maximum seeking, and
instead, actively targets the inclusion boundary, spending its limited budget
on the most contested boundary gaps, where a CE call can actually change which documents are delivered to the generator.

In summary, our main contributions are:

\begin{itemize}
\item \textbf{A New Formulation for RAG Selection:} We frame document selection
  for RAG as a budgeted top-$k$ set-selection problem, and 
  identify the structural mismatch of traditional approaches with this context selection problem.
\item \textbf{State-of-the-Art Diffuse Retrieval:} We propose an efficient, parameter-free Gaussian
Process selector that leverages a MIPS prior to warm-start the relevance
  landscape and utilizes a standardized boundary gap to explicitly target the
  top-$k$ decision threshold. We demonstrate through
  extensive evaluation that BoundaryMORPH consistently outperforms existing active
  selection baselines (BAGEL \citep{kim_bayesian_2026}, RGS
  \citep{xu_beyond_2025}, and single-stage reranking) at equal budgets across
  multiple embedding-CE pairs, achieving the highest selection quality on both
  human relevance judgments (nCG@$k$) and cross-encoder fidelity (CE-nCG@$k$).
\item \textbf{Analysis of Query Multimodality:} We introduce a geometric measure
  of query \emph{multimodality}, revealing that our method's performance advantage
  is fundamentally driven by its ability to navigate complex relevance landscapes.
  Because diffuse queries often exhibit multiple semantic ``peaks'' or
  sub-topics, max-seeking baselines over-measure the primary mode. By explicitly
  avoiding redundant calls on securely top-$k$ documents, BoundaryMORPH naturally
  crosses over to explore and select documents from secondary peaks, yielding
  uniform and growing performance gains on highly multimodal queries.
\end{itemize}

\section{Problem Setting}
\label{sec:setting}

\textbf{Setup.} We define a \emph{diffuse} query to be a query which
requires more than a single or a handful of relevant documents for the best
response. Unlike simple factoid queries that require a single document, diffuse queries (e.g., “Effects of Light Pollution?”) require tens or hundreds. In this work, we focus on the latter.
A query $q$ has an ground truth, unobserved relevance function
$\mathrm{rel}(q, d) \in [0,1]$ over documents $d \in D$ from corpus $D$. Two scoring functions can estimate it.
The \textbf{retriever} $s(q, d)$ is cheap and amortized across queries, and
narrows the corpus to a candidate pool $C$ with $|C| = N$. In practice, $N$ is a
standard first-stage retrieval depth chosen to be large enough to bound maximum
recall. The cross-encoder \textbf{(CE)}, $\fce(q, d) \in [0,1]$, is the more accurate estimate and the
more expensive one: it must be evaluated separately for each query-document pair,
and we may call it at most $B \ll N$ times per query. The consumer of the results
accepts a fixed number $k$ of documents, the \textbf{context capacity}. The system
returns a set $S \subseteq C$ with $|S| = k$.

\textbf{Objective.} We want the returned set to carry as much relevance as
possible,
\begin{equation}
S^{*} = \argmax_{S \subseteq C,\; |S| = k} \sum_{d \in S} \mathrm{rel}(q, d).
\label{eq:objective}
\end{equation}
Given $\mathrm{rel}$, this is trivial: return the $k$ most relevant documents. The
difficulty is that $\mathrm{rel}$ is unknown, that $\fce$ is the closest estimate of
it we can obtain, and that we can afford only $B \ll N$ calls to $\fce$. The problem
is therefore one of resource allocation: which $B$ documents to score, so that the
$k$ we return afterwards are as relevant as possible.

\textbf{Context capacity.} The capacity $k$ is the number of retrieved
documents the RAG generator is given before it begins producing an answer. In
deployed RAG pipelines this is a managed allotment: a finite context window may be
divided among the system prompt, tool schemas, conversation history, retrieved
documents, and response headroom. Hence, the slice reserved for retrieved
documents, $k$, is known, and the task is one of \textbf{selection}: to choose which $k$
documents to return (a set). We view arranging them within the context as a
separate problem \cite{cuconasu_rag_2025}, and in this work, focus on the task of \textit{selecting} the best $k$ documents to fit into the model context. The problem therefore contains two decisions:
\textbf{where} to spend the $B$ CE calls, and \textbf{which} $k$ documents to return once they are
spent.

\section{Method}
\label{sec:morph}

\begin{figure}
    \centering
    \includegraphics{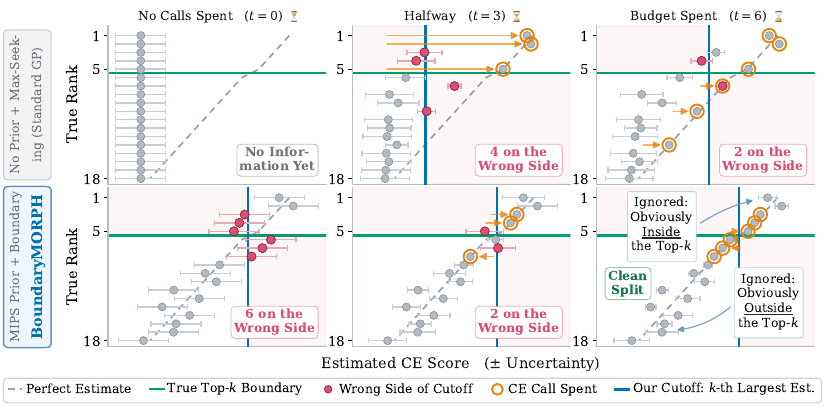}
    \caption{BoundaryMORPH vs.\ a standard GP. \textbf{Top row:} A standard GP (zero prior, max-seeking acquisition) wastes limited budget repeatedly scoring obvious top documents. \textbf{Bottom row:} BoundaryMORPH leverages a MIPS prior and targets the top-$k$ boundary. By ignoring documents safely inside or outside the cutoff, it spends its budget exclusively on resolving contested edge cases.}
    \label{fig:intuition}
\end{figure}

At a high level, our approach uses a Gaussian Process (GP) fitted with specialized
components for this task. We first introduce GPs and the standard way
such an estimator is used to choose what to score. The two subsections that follow
are the specializations fitting it to budgeted selection under a known capacity:
(1) what the estimator starts from, and (2) how it
picks the next document. Full pseudocode is given in Appendix~\ref{sec:app-algorithm}.

\subsection{Preliminaries: Gaussian-process surrogates}
\label{sec:gp-prelim}

\textbf{The estimation problem.} The cross-encoder $\fce$ may be evaluated at $B$ points but must
inform a decision over all $N$. That calls for a \emph{surrogate}: a model that
predicts $\fce(q,d)$ at every unscored document and reports how far each prediction
can be trusted. Uncertainty is necessary, since the selector needs it to tell a
resolved region from an unexamined one.

\textbf{Gaussian processes.} A Gaussian process is a distribution over
functions, $g \sim \mathcal{GP}(\mu_0, K)$, specified by two objects
\citep{williams_gaussian_1995}: a \textbf{prior mean} $\mu_0(d)$, the estimate of
$g(d)$ before any evidence, and a \textbf{kernel} $K(d,d')$, which sets how strongly
an observation at $d'$ constrains the estimate at $d$ and is typically a decreasing
function of the distance between the two documents. It is nonparametric: no weights or
fitting, and all posteriors below are exactly computed. Here $g$ is a generic
function placed under the GP; \S\ref{sec:prior} instantiates it as the residual
of $\fce$ against the MIPS percentile prior.

\textbf{Closed-form conditioning.} After $t$ calls of $g$ on $A_t = \{d_1,\dots,d_t\}$
returning scores $y_t$, the posterior at any document $d$ is Gaussian with mean
and variance
\begin{equation}
\mu_t(d) = \mu_0(d) + \mathbf{k}_t(d)^{\!\top} \widetilde{K}_t^{-1}\big(y_t - \mu_0(A_t)\big),
\qquad
\sigma_t^2(d) = K(d,d) - \mathbf{k}_t(d)^{\!\top} \widetilde{K}_t^{-1}\, \mathbf{k}_t(d),
\label{eq:posterior}
\end{equation}
where $\mathbf{k}_t(d) = \mathbf{k}(d, A_t)$ is the vector of kernel values between $d$ and the
scored documents and $\widetilde{K}_t = K(A_t, A_t) + \sigma_n^2 I$ with
$\sigma_n^2$ a very small observation-noise (jitter) term. Both are available for all
$N$ documents at once, and both update as each new score arrives.

Two properties of these formulas are what selectors can run on. First,
\textbf{uncertainty collapses around evidence}: at a scored document
$\sigma_t^2 \approx \sigma_n^2 \approx 0$, and at its kernel-neighbors $\sigma_t$ is reduced
too, so a region can be recognized as resolved. Second, \textbf{the mean reverts to
the prior far from evidence}: as $\mathbf{k}_t(d) \to 0$ the correction term
vanishes and $\mu_t(d) \to \mu_0(d)$. The posterior is a \emph{local deformation}
of the prior around the points that were paid for.

\textbf{Acquisition: GP-UCB.} Given a surrogate, the standard rule for choosing
the next evaluation is the upper confidence bound
\citep{srinivas_information-theoretic_2012, djolonga_high-dimensional_2013},
\begin{equation}
d^{\star} \;=\; \argmax_{d \,\notin\, A_t} \; \mu_t(d) \,+\, \beta\,\sigma_t(d),
\label{eq:ucb}
\end{equation}
which trades \textbf{exploiting} a high estimated score against \textbf{exploring} an uncertain one,
with the weight $\beta$ setting the balance (commonly tuned or annealed). GP-UCB is formulated and analyzed as a method for locating the
\textbf{global maximum} of $g$ with few function evaluations. A GP surrogate thus leaves exactly two things to specify: the pair $(\mu_0, K)$,
and the acquisition rule.

\subsection{The MIPS percentile prior}
\label{sec:prior}

\textbf{MIPS prior.} While the MIPS ranking may not be as accurate as CE relevance, it is certainly
not noise, and it is free (pre-computed). It encodes mostly-reliable geometry that may be
miscalibrated and locally wrong. To take advantage of this, we have the GP model
the \textbf{residual}, $r(d) = \fce(q,d) - \mu_0(d)$, rather than the raw CE field. That
is, we aim to approximate where, and how much to \textbf{morph} the embedding
similarity in order to reach CE relevance. Because the CE outputs are calibrated
probabilities in $[0,1]$, we can meaningfully model the residual against a $[0,1]$
MIPS percentile rank. We set $\mu_0(d)$ to the document's \textbf{MIPS percentile rank} within $C$
($\approx 1$ at the head, $\approx 0$ in the tail), which is scale-free, requiring no
per-query calibration. The posterior mean is then $\mu_0$ plus kernel-weighted observed residuals. Far
from any observation $\mu_t \to \mu_0$: the surrogate falls back to trusting the
MIPS's ranking. Calls \emph{bend} the MIPS's surface; they never rebuild it. This
is also the easier estimation problem: the residual is near zero across most of
the pool, so ``no correction'' is a good default almost everywhere and the GP only
has to represent the surprises.

\textbf{Kernel and query observation.} We use a directional kernel (vMF) on unit embeddings,
$K(d,d') = \exp\big(\kappa\,(\cos(d,d') - 1)\big)$
\citep{mukhopadhyay_estimating_2020, mardia_directional_2009}. Alternative kernels are compared in Appendix~\ref{sec:app-kernel}.
Residuals are Z-standardized before conditioning and de-standardized after.  Similar to \citet{kim_bayesian_2026}, we enter the query vector into the GP before any budget is spent, as a zero-cost
observation at maximal relevance ($y = 1$), as if we had observed a maximally relevant document. This costs no CE call and gives the surrogate one anchor of known value from the start, and pinches the variance near the query.

\subsection{The boundary acquisition rule}
\label{sec:acq}

\textbf{The mismatch of maximum-seeking acquisition.} The objective from
\S\ref{sec:setting} targets \emph{membership} in a set of size $k$. We argue that standard
maximum-seeking acquisition (e.g., GP-UCB) is fundamentally misaligned with this goal.
First, it wastes calls on documents already securely inside the top-$k$. Their exact
\emph{value} is uncertain, but their \emph{membership} is not; they will enter the
context window whether or not we pay to measure them. Symmetrically, it explores regions
securely below the cutoff, confirming exclusions that were never in doubt. A maximum-seeker
has no concept of the known capacity $k$, so it cannot ignore documents whose fate is already
settled. Instead, a useful CE call is one that could plausibly \emph{move a document across}
the boundary, as only these calls actually change the final context fed to the LLM.

\textbf{Relation to active set ordering.} Our selection rule is adapted from
Mean Prediction in \cite{nguyen_active_2024} and, in the independent-arm
case, from LUCB \citep{kalyanakrishnan_pac_nodate}, which estimates the
\textbf{top-$k$ set} of a blackbox function rather than its maximizer, and spends
its evaluations on the boundary comparisons the posterior leaves ambiguous.
Notably, standard GP-UCB is the $k = 1$ special case: a
max-seeking selector is inherently minimizing regret of the global maximum.

\textbf{Rule.} At each step, we partition $C$ by the current posterior mean $\mu_t$ into its \textbf{top-$k$}, $S_t$ (the \emph{incumbents}, our running guess at the returned set $S$), and the rest, $C \setminus S_t$ (the \emph{challengers}). See Figure~\ref{fig:intuition}'s purple line. Membership is in genuine doubt only at the boundary: an incumbent whose lower confidence bound dips below a challenger risks being wrongly retained, and a
challenger whose upper confidence bound rises above an incumbent risks being
wrongly dropped. The contested pairs are those close in estimated score
\emph{relative to how uncertain we are about them}, so we select the pair with the
smallest \textbf{standardized boundary gap},
\begin{equation}
(j^*, i^*) \;=\; \argmin_{j \in S_t,\; i \in C \setminus S_t} \; \frac{\mu_t(j) - \mu_t(i)}{\sigma_t(j) + \sigma_t(i)},
\label{eq:argmin}
\end{equation}
the pair most likely to sit on the wrong sides of the cutoff under the current estimates. We then spend the CE call on whichever of the two has
\textbf{higher $\sigma$} (the outcome we can least predict). Equivalently, $j^{*}$
is the most-threatened incumbent and $i^{*}$ the strongest challenger.

\textbf{Complexity.} The kernel is built once, $O(N^2)$: one entry per document pair at test time. Updating the posterior requires $O(B^3 N)$, due to the matrix inversion, but we find in Appendix~\ref{sec:latency} that $O(B)$ CE inference dominates the method's latency. Full derivation in Appendix~\ref{sec:app-complexity}.
\section{Experimental Setup}
\label{sec:setup}

\subsection{Datasets}
\label{sec:datasets}


\textbf{Datasets.} We evaluate on three datasets featuring broad, open-ended queries over news, government, and encyclopedic corpora: NeuCLIRBench-monolingual-eng \citep{lawrie_neuclirbench_2025}, Robust04 \citep{voorhees_trec_2005}, and TravelDest \citep{wen_elaborative_2024}, with relevance labels from \citet{kim_bayesian_2026}. To ensure diffuseness, we filter out the minority of queries with $<20$ relevant documents, leaving 85, 187, and 99 queries, respectively, with median relevant document counts of \textbf{51 / 62 / 462}. To make the setting tractable and similar to realistic use-cases, we restrict evaluation to a per-query candidate set: the top $N=10,000$ by Qwen3-Embedding-4B MIPS \citep{zhang_qwen3_2025}. Full statistics are in Appendix~\ref{sec:app-datasets}.

\subsection{Metrics}
\label{sec:metrics}

\textbf{Using nCG@$k$ and CE-nCG@$k$.} We use normalized cumulative gain (\textbf{nCG@$k$}) against human relevance judgments
as our primary metric \citep{jarvelin_cumulated_2002}. Intuitively, this metric measures, \textit{``compared to the \textbf{best possible set} of $k$ documents that could have been retrieved, how relevant was the set that \textbf{was} retrieved?''}. We also report \textbf{CE-nCG@$k$}, the same quantity computed against the CE's scores rather than human judgments. Concretely:

\begin{equation}
\mathrm{nCG@}k = \frac{\sum_{d \in S}\mathrm{rel}(q,d)}{\sum_{d \in S^{*}}\mathrm{rel}(q,d)}, \qquad
\mathrm{CE\text{-}nCG@}k = \frac{\sum_{d \in S}\fce(q,d)}{\sum_{d \in S^{\dagger}}\fce(q,d)},
\label{eq:ncg}
\end{equation}

where $S$ is the selected set of size $k$, $S^{*}$ is the optimal set from Eq.~\ref{eq:objective}, and $S^{\dagger}$ is the analogous optimal set of size $k$ under the CE's own scores. Hence, \textbf{nCG@$k$} is the count of relevant documents
returned, divided by the number a perfect selector would fit into $k$ slots. It
reaches 1 exactly when the returned set is as relevant as the most relevant possible set. Reporting both metrics separates two claims: CE-nCG measures how well the budget is allocated to approximate $\fce$, while nCG@$k$ measures the retrieval quality against human judgments, rather than only tracking the CE more faithfully.

\textbf{Why not recall or nDCG@$k$.} Because relevant documents often exceed $k$, nCG@$k$ directly asks: \emph{of the $k$ documents the generator can read, how many are
relevant?} Metrics such as nDCG and MAP score the
\emph{ordering} within the returned set, which are not well-suited for measuring which $k$ documents reach the generator's context \citep{trappolini_redefining_2025, samuel_beyond_2026}.

\todo{this Metrics section is way too long right now}

\subsection{Methods}
\label{sec:baselines}

The dual-encoder (embedding model) and CE are \textbf{multilingual-e5-large}
\citep{wang_multilingual_2024} and \textbf{zerank-2} \citep{pipitone_zelo_2025} respectively, with other models in Appendix~\ref{sec:app-robustness}, and implementation details in ~\ref{sec:app-gp}). \textbf{BM25} \citep{robertson_okapi_1994} is a lexical floor: applied as a re-ranking of the
shared top-10k pool and truncated at $k$, giving one fixed ordering per query.
\textbf{Single-Stage Rerank (SS)} is the standard policy: score the MIPS
top-$B$ with the CE, place them above the tail, and leave the tail in MIPS
order.
\textbf{BAGEL} \citep{kim_bayesian_2026} is a previous GP approach using
zero-prior, and standard UCB over the whole pool with an explore/exploit term which we sweep.
\textbf{RGS} \citep{xu_beyond_2025} is a greedy kNN graph-based approach. Because it lacks a way to rank documents it never scored, scored documents are placed as a band above the
unscored tail, which is left in MIPS order.
RGS and BAGEL's hyperparameters are each tuned and selected on the evaluation metric per budget. Full implementation and sweep details for every baseline are in Appendix~\ref{sec:app-baselines}.

\section{Results and Analysis}
\label{sec:results}

\begin{table}
\centering
\fontsize{9}{9}\selectfont
\setlength{\tabcolsep}{5pt}
\caption{Top-$k$ relevance (\%) under CE budget $B$ for various methods. Best per column in \textbf{bold}, second-best \underline{underlined}. BoundaryMORPH outperforms baselines, with BAGEL a clear second, due to the ability to estimate relevance of unscored documents.}
\label{tab:main-A}
\begin{tabular}{c l rrr rrr !{\hskip 0.4em\vrule\hskip 0.4em} rrr rrr}
\toprule
 &  & \multicolumn{6}{c}{nCG@$k$ (\%, human judgments)} & \multicolumn{6}{c}{CE-nCG@$k$ (\%)} \\
\cmidrule(lr){3-8}\cmidrule(lr){9-14}
 &  & \multicolumn{3}{c}{$B$=25} & \multicolumn{3}{c}{$B$=50} & \multicolumn{3}{c}{$B$=25} & \multicolumn{3}{c}{$B$=50} \\
\cmidrule(lr){3-5}\cmidrule(lr){6-8}\cmidrule(lr){9-11}\cmidrule(lr){12-14}
 & Method\hspace{3pt}\rotatebox[origin=lB]{58}{\rule{1em}{0.8pt}}\hspace{3pt}$k$ & 50 & 100 & 200 & 50 & 100 & 200 & 50 & 100 & 200 & 50 & 100 & 200 \\
\midrule
\multirow{5}{*}{\rotatebox[origin=c]{90}{NeuCLIR}} & BM25 & 38.7 & 46.8 & 57.6 & 38.7 & 46.8 & 57.6 & 69.8 & 67.3 & 66.5 & 69.8 & 67.3 & 66.5 \\
 & SS & 39.6 & 46.2 & 55.4 & 39.6 & 46.2 & 55.4 & 73.1 & 70.0 & 68.5 & 73.1 & 70.0 & 68.5 \\
 & RGS & 39.1 & 46.9 & 55.7 & 37.0 & 47.5 & 56.6 & 72.3 & 70.3 & \underline{68.7} & 68.8 & 70.3 & 69.1 \\
 & BAGEL & \underline{47.3} & \underline{51.4} & \underline{58.5} & \underline{52.9} & \underline{55.7} & \underline{62.3} & \underline{79.7} & \underline{71.5} & 66.9 & \underline{86.9} & \underline{77.6} & \underline{71.6} \\
 & \textbf{MORPH} & \textbf{49.8} & \textbf{55.1} & \textbf{62.7} & \textbf{54.7} & \textbf{59.2} & \textbf{65.1} & \textbf{85.7} & \textbf{80.9} & \textbf{77.3} & \textbf{90.5} & \textbf{85.7} & \textbf{81.9} \\
\midrule
\multirow{5}{*}{\rotatebox[origin=c]{90}{Robust04}} & BM25 & 34.7 & 40.0 & \underline{49.7} & 34.7 & 40.0 & 49.7 & 49.7 & 47.6 & 47.8 & 49.7 & 47.6 & 47.8 \\
 & SS & 34.4 & 37.1 & 44.8 & 34.4 & 37.1 & 44.8 & 54.5 & 51.5 & 51.0 & 54.5 & 51.5 & 51.0 \\
 & RGS & 35.0 & 38.1 & 45.7 & 34.2 & 38.8 & 46.6 & 54.8 & 52.0 & \underline{51.4} & 53.5 & 52.3 & 51.8 \\
 & BAGEL & \underline{40.3} & \underline{41.0} & 46.1 & \underline{48.7} & \underline{48.0} & \underline{53.0} & \underline{59.9} & \underline{53.2} & 50.2 & \underline{68.4} & \underline{59.6} & \underline{55.2} \\
 & \textbf{MORPH} & \textbf{45.2} & \textbf{46.4} & \textbf{52.1} & \textbf{52.3} & \textbf{51.5} & \textbf{57.1} & \textbf{66.0} & \textbf{60.5} & \textbf{57.3} & \textbf{72.8} & \textbf{65.8} & \textbf{61.6} \\
\midrule
\multirow{5}{*}{\rotatebox[origin=c]{90}{TravelDest}} & BM25 & 23.0 & 21.9 & 23.0 & 23.0 & 21.9 & 23.0 & 57.3 & 57.4 & 57.7 & 57.3 & 57.4 & 57.7 \\
 & SS & 30.9 & 28.4 & 27.9 & 30.9 & 28.4 & 27.9 & 64.2 & 64.5 & 64.8 & 64.2 & 64.5 & 64.8 \\
 & RGS & 32.3 & 29.4 & 28.6 & 32.9 & 30.4 & 29.3 & 63.9 & 64.5 & 65.0 & 64.4 & 65.3 & 65.5 \\
 & BAGEL & \underline{40.9} & \underline{35.5} & \underline{33.7} & \underline{46.6} & \underline{40.7} & \underline{37.5} & \underline{75.9} & \underline{72.4} & \underline{69.9} & \underline{82.0} & \underline{77.5} & \underline{74.5} \\
 & \textbf{MORPH} & \textbf{45.1} & \textbf{39.8} & \textbf{37.8} & \textbf{48.9} & \textbf{43.1} & \textbf{40.8} & \textbf{81.1} & \textbf{78.4} & \textbf{76.5} & \textbf{86.5} & \textbf{82.9} & \textbf{80.3} \\
\bottomrule
\end{tabular}
\end{table}

The evaluation addresses four questions in turn: (\S\ref{sec:main-results}) Does BoundaryMORPH select better sets than existing selectors at equal budget? (\S\ref{sec:ablations-analysis}) Are both of its components necessary (MIPS prior/boundary acquisition), and robust to the capacity $k$? (\S\ref{sec:multimodality}) How does BoundaryMORPH achieve its improvement, and which queries, geometrically speaking, yield the greatest advantage? In Appendix~\ref{sec:latency}, we also compare wall-clock latency, including against an agentic retrieval loop.

\subsection{Main Results}
\label{sec:main-results}

\textbf{BoundaryMORPH outperforms baselines in all configurations.} Table~\ref{tab:main-A} shows the main experimental results. BoundaryMORPH outperforms baselines in all (dataset $\times$ $B$ $\times$ $k$) combinations, with the biggest gains being at higher $k$ values (average of +3.8 pp nCG@100 and +7.3 pp CE-nCG@200 over second-best baseline). BAGEL outperforms the other baselines, because it is the only baseline which also can use CE calls to update the relevance of \textit{unscored} documents. In Appendix~\ref{sec:app-robustness}, we show BoundaryMORPH's performance gap holds across choice of embedding and CE model. We also see that \textbf{CE gains carry over to human judgments.} The trend persists across the two metrics, indicating that better
approximation of the CE's own top-$k$ translates into more human-judged-relevant
documents delivered.

\subsection{Ablations and Sensitivity}
\label{sec:ablations-analysis}

\begin{wraptable}{R}{0.5\textwidth}
\vspace{\dimexpr-\intextsep-48pt\relax}
\setlength{\abovecaptionskip}{0pt}
\centering
\footnotesize
\setlength{\tabcolsep}{3.5pt}
\caption{Ablations: nCG@$k$ (\%) averaged over datasets and embedder$\times$CE pairs; indented rows give the pp change.}
\label{tab:ablation}
\vspace{3pt}
\setlength{\aboverulesep}{0pt}
\setlength{\belowrulesep}{1pt}
\begin{tabular}{l rrr rrr}
\toprule
 & \multicolumn{3}{c}{$B$=25} & \multicolumn{3}{c}{$B$=50} \\[-1.5pt]
\cmidrule(lr){2-4}\cmidrule(lr){5-7}
\multicolumn{1}{r}{$k$} & 50 & 100 & 200 & 50 & 100 & 200 \\[-1.5pt]
\midrule
\textbf{MORPH} & 47.1 & 47.9 & 52.6 & 51.3 & 51.3 & 54.9 \\
\quad w/o acq. & \cellcolor[rgb]{1.000,0.631,0.631} -2.6 & \cellcolor[rgb]{1.000,0.648,0.648} -2.5 & \cellcolor[rgb]{1.000,0.704,0.704} -2.1 & \cellcolor[rgb]{1.000,0.783,0.783} -1.5 & \cellcolor[rgb]{1.000,0.673,0.673} -2.3 & \cellcolor[rgb]{1.000,0.778,0.778} -1.5 \\
\quad w/o prior & \cellcolor[rgb]{1.000,0.569,0.569} -3.0 & \cellcolor[rgb]{1.000,0.553,0.553} -3.1 & \cellcolor[rgb]{1.000,0.450,0.450} -3.8 & \cellcolor[rgb]{1.000,0.694,0.694} -2.1 & \cellcolor[rgb]{1.000,0.675,0.675} -2.3 & \cellcolor[rgb]{1.000,0.662,0.662} -2.3 \\
\bottomrule
\end{tabular}
\end{wraptable}

\textbf{Both the MIPS prior and the boundary rule are necessary.}\label{sec:ablation} To test the contribution of BoundaryMORPH's two components, we perform two isolated removals, averaged over the four model pairs and three datasets: \textbf{without
acquisition}, the boundary acquisition rule reverts to standard UCB at its own best swept $\beta$; \textbf{without prior}, the MIPS percentile prior mean becomes a zero mean.

As seen from Table~\ref{tab:ablation}, removing boundary acquisition lowers nCG@$k$ by 1.5--2.6 percentage points, and removing the prior by 2.1--3.8. The drop holds in every per-dataset and CE-nCG cell as well (Appendix~\ref{sec:app-ablation-full}), which verifies the utility of each component separately.

\begin{figure}
    \centering
    \begin{minipage}[t]{0.6\textwidth}
        \centering
        \includegraphics[width=3.2in]{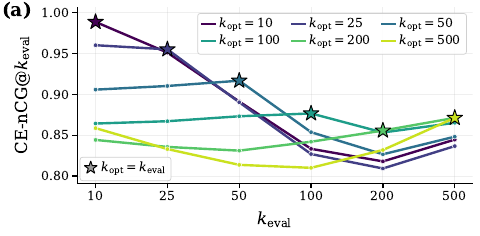}
    \end{minipage}%
    \begin{minipage}[t]{0.4\textwidth}
        \centering
        \includegraphics[width=2.1in]{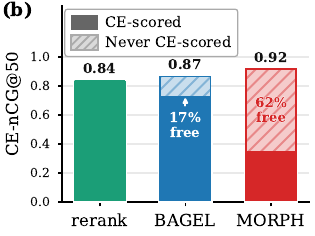}
    \end{minipage}
    \par\vspace{6pt}
    \includegraphics{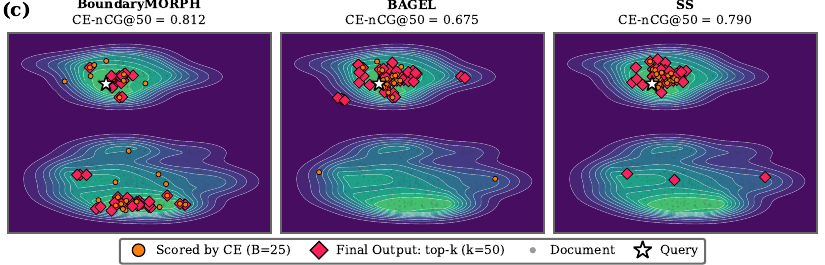}
    \caption{\ul{\textbf{\textit{(a)}}}: \textbf{the acquisition rule optimizes for the exact capacity it is
    given.} We plot CE-nCG at different \textit{evaluated} capacities $k_{\mathrm{eval}}$ ($x$-axis), with one curve per \textit{optimized} capacity $k_{\mathrm{opt}}$ (the one given to the algorithm); when evaluated at the optimized $k$, $k_{\mathrm{eval}}$=$k_{\mathrm{opt}}$ (stars), performance peaks there.
    \ul{\textbf{\textit{(b)}}}: \textbf{BoundaryMORPH wins \emph{because} it does not spend its budget
    measuring the obvious top.} Share of each method's top-$k$ CE-nCG total coming 
    from documents it never scored (``free''), at $B$=50, $k$=50.
    \ul{\textbf{\textit{(c)}}}: \textbf{Skipping the obvious top buys the budget to reach a second
    mode.} One query's pool, plotted via t-SNE over document embeddings and
    contoured by CE relevance, with each method's \emph{scored} and finally
    \emph{selected} documents marked. Because BoundaryMORPH does not score
    documents obviously inside the top-$k$, it has budget left to
    explore, find a second mode, and exploit it, while
    max-seeking baselines concentrate calls on the primary mode.}
    \label{fig:k-opt-vs-k-eval}
    \label{fig:scored-vs-selected}
    \label{fig:contour-scored-topk}
\end{figure}

\textbf{BoundaryMORPH exploits its knowledge of the context capacity.}\label{sec:ksens} To test how the input context capacity $k$ impacts our algorithm, we run BoundaryMORPH
with varying values of $k$ (the optimized value, $k_{\mathrm{opt}}$). However, instead of evaluating on the corresponding CE-nCG@$k_{\mathrm{opt}}$, we
evaluate at \emph{many} capacity values $k_{\mathrm{eval}}$, to see what happens
when we evaluate at a \emph{different} capacity from what we gave our method. As seen in Figure~\ref{fig:k-opt-vs-k-eval}(a), for nearly all evaluated $k_{\mathrm{eval}}$,
the run that was given $k_{\mathrm{eval}} = k_{\mathrm{opt}}$ is the best of all runs. That is, BoundaryMORPH truly is optimizing for $k_{\mathrm{opt}}$, and fundamentally exploits the final evaluation's disregard for order within and outside the top-$k$.

\subsection{Sources of Improvement}
\label{sec:multimodality}

\textbf{Budget is spent at the boundary rather than on secure incumbents.}\label{sec:budget-alloc} To investigate how BoundaryMORPH achieves performance gains, we analyze how each method allocates its budget. Figure~\ref{fig:scored-vs-selected}(b) reports for 3 methods: among the sum relevance in the final top-$k$, how much relevance came from \textit{CE-scored} documents vs. \textit{unscored}. We see that BoundaryMORPH scores far \textit{fewer} of the documents it finally returns. It declines to
spend calls on high-confidence, high-posterior documents whose membership is not in
doubt, freeing budget for the boundary where a call can flip the answer, whereas BAGEL spends far more calls at the top: seeking a global maximum. Purely reranking with the CE by definition can only include documents that are scored. Figure~\ref{fig:contour-scored-topk}(c)
shows one NeuCLIR query's \textit{scored} and \textit{finally selected} documents on its relevance contour: by avoiding over-scoring the main mode, BoundaryMORPH is able to cross over and
explore a secondary mode.

\begin{figure}[t]
    \centering
    \includegraphics{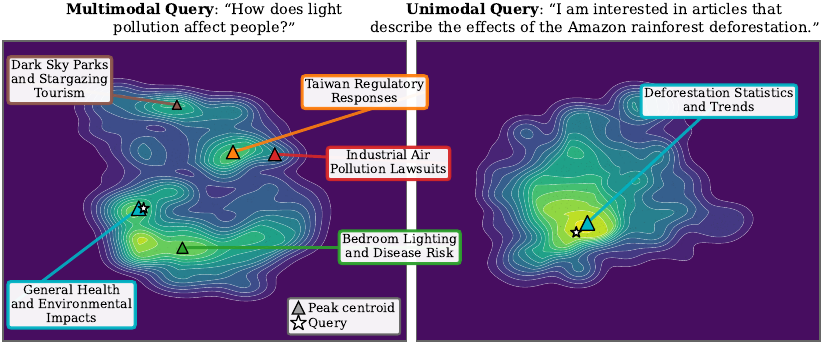}
    \caption{\textbf{Geometric peaks are genuinely distinct sub-topics.} CE
    relevance contours over t-SNE embedding space for a \textit{multimodal} and a \textit{unimodal} query,
    with the extracted peaks and LLM-generated peak labels. The multimodal query retains 5 peaks, each a distinct
    facet of the information need rather than one blurred high-relevance region;
    the unimodal query (Amazon deforestation) retains one. Peak extraction
    details are in Appendix~\ref{sec:app-peaks}.}
    \label{fig:qualitative-peaks}
\end{figure}

\textbf{The advantage increases with query multimodality.} Figure~\ref{fig:contour-scored-topk}(c) suggests the advantage comes from reaching secondary relevance modes, so we measure how \emph{multimodal} each query is: whether its relevant documents form one tight group or several distinct ones. Treating CE score as height over an embedding kNN graph, we extract relevance \emph{peaks} with persistence-based clustering \citep{chazal_persistence-based_2013} and define a query's \textit{multimodal mass} as the prominence-weighted relevance outside its tallest peak (Appendix~\ref{sec:app-peaks}). Figure~\ref{fig:qualitative-peaks} contrasts a multimodal query (\textit{light pollution}), whose peaks are distinct sub-topics, with a unimodal one (\textit{Amazon deforestation}). On NeuCLIR, multimodal mass predicts per-query advantage over BAGEL (Spearman $\boldsymbol{\rho = +0.59}$, $p = 3.5\times 10^{-9}$): \textbf{BoundaryMORPH gains most where it must reach ``across'' valleys to separate modes.} An LLM audit confirms that peaks are semantically coherent groups of documents (Appendix~\ref{sec:app-peaks}).

\section{Related Work}
\label{sec:related}

\textbf{Budgeted reranking.} \emph{Adaptive} retrieval dynamically chooses what to score at query time. GAR \citep{macavaney_adaptive_2022}, Quam \citep{rathee_quam_2025}, SlideGar \citep{rathee_guiding_2025}, and RGS \citep{xu_beyond_2025} iteratively expand graph neighborhoods of high-scoring candidates, while budget-aware cascades \citep{rashid_ecorank_2024, li_evirerank_2026} and list truncation \citep{meng_ranked_2024} dynamically halt reranking over a fixed order. Crucially, all are value-seeking and cannot estimate an unscored document. Closest in intent is AcuRank \citep{yoon_acurank_2025}, which maintains Bayesian beliefs to resolve top-$k$ membership doubt. However, its beliefs update independently, meaning an unscored document's posterior never updates based on a scored neighbor's. Methodologically closest is BAGEL \citep{kim_bayesian_2026}, a pool GP differing via deterministic warm-starting, a zero prior mean, and a maximum-seeking acquisition (GP-UCB).

\textbf{Budgeted Selection Goals.} Level-set estimation samples where confidence intervals straddle a known threshold \citep{bryan_active_2005, gotovos_active_2013}. Relatedly, active set ordering \citep{nguyen_active_2024} and LUCB-style bandits over independent arms \citep{kalyanakrishnan_pac_2012, jiang_practical_2017, pony_col-bandit_2026} resolve top-$k$ membership for a known $k$. We observe that budgeted retrieval is fundamentally an active set ordering problem.

\textbf{Indexing-time methods and agentic retrieval.} Orthogonal offline methods optimize the corpus before querying via document expansion \citep{nogueira_document_2019}, reranker distillation \citep{schlatt_rank-distillm_2025}, or LLM-built graphs \citep{sarthi_raptor_2024, edge_local_2025, gutierrez_hipporag_2024}. These change \emph{what the pool contains}; we optimize which members reach the generator. Another orthogonal family uses LLMs to interleave search and generation \citep{khattab_demonstrate-search-predict_2023, khattab_dspy_2024, jiang_active_2023, jeong_adaptive-rag_2024}, sometimes via reinforcement learning \citep{jin_search-r1_2025}, which forms the basis of deep-research agents \citep{huang_deep_2025, coelho_deepresearchgym_2025}. Because these systems treat the retriever as a standalone tool, our budgeted selector can trivially plug into their loops to raise accuracy and cut search calls \citep{sharifymoghaddam_rerank_2026, meng_revisiting_2026}.

\section{Conclusion}
\label{sec:conclusion}

We introduced BoundaryMORPH, a budget-aware document selection method for diffuse Retrieval-Augmented Generation. Rather than wasting limited cross-encoder calls on documents securely within the context capacity, BoundaryMORPH uses a Gaussian Process with a MIPS prior to explicitly target the top-$k$ decision boundary. This approach achieves state-of-the-art selection quality over existing maximum-seeking baselines. By avoiding redundant measurements of primary topics, it successfully navigates multimodal queries to capture diverse semantic peaks, making it exceptionally well-suited for complex, deep-research applications. While BoundaryMORPH offers significant advantages for diffuse retrieval, we acknowledge certain system and algorithmic limitations. Specifically, the method currently relies on sequential execution, assumes a discrete document capacity $k$ rather than strict token limits, and scales cubically with extreme budgets. We discuss these limitations and potential future directions in detail in Appendix~\ref{sec:limitations}. \todo{I want this last bit OK'ed by mentors.}


\subsection*{AI use statement}
In this work, we used generative AI tools for polishing writing, including shortening and feedback. All text was manually written or reviewed. Additionally, AI tools were used for literature discovery and summarization. All cited works were manually curated with Zotero, and checked for the individual claims for which they were cited. AI tools were also used for coding, but artifacts were manually verified and tested. We have reviewed all AI-assisted work. We take responsibility for the final content of this work, including text, claims or artifacts produced with the aid of generative AI.

\subsection*{Reproducibility statement}
Appendix~\ref{sec:app-method} contains all details necessary for reproducing our method (including pseudocode), while Appendix~\ref{sec:app-setup} contains all details necessary to replicate our experiments, including dataset processing details, baseline details, and prompts. We also include with this submission an anonymized codebase implementing our method and the baselines.


\bibliography{iclr2026_conference}
\bibliographystyle{iclr2027_conference}

\appendix

\section{Limitations}
\label{sec:limitations}

While BoundaryMORPH successfully optimizes retrieval for diffuse queries under budget constraints, we identify several limitations and system assumptions that offer directions for future work:

\textbf{Sequential execution and hardware utilization.} Because BoundaryMORPH updates the GP posterior after each cross-encoder call, it fundamentally requires sequential execution. As discussed in Appendix~\ref{sec:latency}, while this is competitive on a single GPU, heavily parallelized production environments can amortize single-stage reranking via massive batching. Extending our boundary acquisition rule to select batches of documents simultaneously (e.g., auto-kriging) could improve hardware utilization. In addition, many applications in practice are willing to trade off some latency for higher quality retrieval (e.g., long deep research loops).

\textbf{GP scaling at massive budgets.} The exact GP posterior update scales cubically with the budget and linearly with the pool size ($O(B^3 N)$). While our latency analysis (Appendix \ref{sec:app-latency}) shows the algorithm is lightweight for realistic RAG budgets ($B \le 200$), the matrix inversion bottleneck becomes significant at extreme budgets (e.g., $>30$s at $B=500$). If CE inference becomes drastically cheaper in the future, the relative overhead of the GP will increase. Future work could explore algorithmic improvements, e.g. using rank-1 matrix updates.

\textbf{Document counts vs. token limits.} We formulate the RAG capacity as a strict document count $k$, whereas real-world LLM context windows are strictly bounded by \emph{tokens}. However, because modern RAG pipelines heavily rely on text chunking prior to indexing, candidate documents typically fall into a relatively narrow and predictable token band. This system practice is what allows the discrete document count $k$ to serve as a highly realistic and practical proxy for context capacity. Nonetheless, adapting the boundary gap for RAG systems that utilize variable-length chunks remains an interesting future challenge.

\textbf{Scope of queries.} As we explicitly evaluate on \emph{diffuse} queries requiring many sources, our method is tailored to exploratory search and deep-research agents. For simple factoid queries requiring only a single, specific document ($k=1$), the exploratory benefits of BoundaryMORPH do not apply, and standard maximum-seeking approaches will suffice. 

\textbf{Selection versus within-context ordering.} As defined in \S\ref{sec:setting}, BoundaryMORPH focuses purely on set \emph{selection} (which $k$ documents are retrieved). Because we actively avoid spending budget on the secure top-$k$ (Figure \ref{fig:scored-vs-selected}(b)), the exact relative ordering of those unscored documents relies entirely on the GP estimation. However, we argue that \emph{whether} a document is included in the RAG context is fundamentally more important than \emph{where} it is placed. Recent research has demonstrated that even with perfect relevance scores, sophisticated context-rearrangement strategies often fail to outperform random shuffling \citep{cuconasu_rag_2025}. Therefore, determining the optimal arrangement of context is a distinct, largely open research question, and we view maximizing the sheer volume of relevant information delivered to the context as the primary bottleneck for diffuse queries.


\section{Method Details}
\label{sec:app-method}

\subsection{Algorithm pseudocode}
\label{sec:app-algorithm}

Algorithms~\ref{alg:posterior}--\ref{alg:boundarymorph} give full pseudocode for the procedure described in \S\ref{sec:acq}. As in \S\ref{sec:prior}, the GP is fit not to the raw CE score but to the \textbf{residual} against the MIPS prior, and that residual is Z-standardized before conditioning and de-standardized after (Algorithm~\ref{alg:posterior}); Algorithm~\ref{alg:boundarymorph} is the outer acquisition loop that calls it.

\begin{algorithm}[h]
\caption{\textsc{Posterior}: residual GP conditioning}
\label{alg:posterior}
\begin{algorithmic}[1]
\Function{Posterior}{observed set $A$, observed scores $\mathbf{y}$, prior mean $\mu_0$, kernel $K$, pool $C$}
    \State $\mathbf{e} \gets \mathbf{y} - \mu_0(A)$ \Comment{residuals of the observations, $r(d) = \fce(q,d) - \mu_0(d)$, \S\ref{sec:prior}}
    \State $\mathbf{z} \gets \big(\mathbf{e} - \mathrm{mean}(\mathbf{e})\big) / \mathrm{std}(\mathbf{e})$ \Comment{Z-standardize the residuals}
    \State $(\tilde\mu, \tilde\sigma^2) \gets$ zero-mean GP posterior of $K$ given $(A, \mathbf{z})$, for all $d \in C$ \Comment{Eq.~\ref{eq:posterior} with $\mu_0 \equiv 0$}
    \State $\mu(d) \gets \mu_0(d) + \mathrm{mean}(\mathbf{e}) + \mathrm{std}(\mathbf{e}) \cdot \tilde\mu(d)$ for all $d \in C$ \Comment{de-standardize}
    \State $\sigma^2(d) \gets \mathrm{std}(\mathbf{e})^2 \cdot \tilde\sigma^2(d)$ for all $d \in C$
    \State \Return $(\mu, \sigma^2)$
\EndFunction
\end{algorithmic}
\end{algorithm}

\begin{algorithm}[h]
\caption{BoundaryMORPH}
\label{alg:boundarymorph}
\begin{algorithmic}[1]
\Require query $q$; retriever $s$; cross-encoder $\fce$; budget $B$; capacity $k$
\Ensure selected set $S \subseteq C$, $|S| = k$
\State $C \gets$ top-$N$ candidates from $s(q)$
\State Build kernel $K$ over $C \cup \{q\}$ (\S\ref{sec:prior}); set prior mean $\mu_0(d) \gets$ MIPS percentile rank of $d$ in $C$
\State $A \gets \{q\}$, $\mathbf{y} \gets \{1\}$ \Comment{query enters as a zero-cost observation at $y=1$}
\State $(\mu, \sigma^2) \gets \Call{Posterior}{A, \mathbf{y}, \mu_0, K, C}$
\For{$t = 1, \dots, B$}
    \State $S_t \gets$ top-$k$ of $C$ by $\mu$ \Comment{incumbents; challengers are $C \setminus S_t$}
    \State $(j^*, i^*) \gets \argmin_{j \in S_t,\, i \in C \setminus S_t} \frac{\mu(j) - \mu(i)}{\sigma(j) + \sigma(i)}$ \Comment{standardized boundary gap, Eq.~\ref{eq:argmin}}
    \State $d \gets \argmax\big(\sigma(j^*), \sigma(i^*)\big)$ \Comment{score whichever of the pair is less certain}
    \State $y \gets \fce(q, d)$ \Comment{one CE call}
    \State $A \gets A \cup \{d\}$, $\mathbf{y} \gets \mathbf{y} \cup \{y\}$
    \State $(\mu, \sigma^2) \gets \Call{Posterior}{A, \mathbf{y}, \mu_0, K, C}$
\EndFor
\State \Return the top-$k$ documents of $C$ ranked by $\mu$ (scored documents at their true CE score, unscored documents at $\mu$)
\end{algorithmic}
\end{algorithm}

\subsection{Time complexity}
\label{sec:app-complexity}

This section expands on the costs stated in \S\ref{sec:acq}'s \textbf{Complexity} paragraph. The kernel $K$ is a Gram matrix over the pool (augmented with the query row of \S\ref{sec:prior}), one entry per pair of documents, so building it once before any CE call is spent costs $O(N^2)$.

Each of the $B$ steps of Algorithm~\ref{alg:boundarymorph}'s loop re-conditions the GP on a growing observed set $A_t$, $|A_t| = t$. Rather than refactorizing from scratch, the Cholesky factor of the $t \times t$ observed-block kernel is updated incrementally at $O(t^2)$, but reading off the posterior mean and variance at \emph{every} one of the $N$ pool documents requires a triangular solve against that factor per document, an $O(t^2)$ solve repeated $N$ times, i.e.\ $O(t^2 N)$ at step $t$. Summing over the run, $\sum_{t=1}^{B} O(t^2 N) = O(B^3 N)$, which is the matrix-inversion cost the main text refers to. The acquisition rule (Eq.~\ref{eq:argmin}) is comparatively cheap: it evaluates the standardized boundary gap over the full incumbent$\times$challenger cross product, $O(k(N-k))$ per step and $O(Bk(N-k))$ over the run. Because $B, k \ll N$ in practice, both terms are far below their worst case, which is why Appendix~\ref{sec:latency} finds that the $O(B)$ sequential CE calls, not the $O(B^3N)$ GP bookkeeping, dominate wall-clock latency at the budgets used in this paper.

\subsection{GP implementation}
\label{sec:app-gp}

The kernel is the vMF kernel $K(d,d') = \exp(\kappa(\cos(d,d') - 1))$ on
unit-normalized embeddings, with the diagonal forced to $1$. Its bandwidth is set
per query by the median heuristic,
$\kappa = \log 2 / (1 - \mathrm{median}_{i<j} \cos(d_i,d_j))$ over the pool,
falling back to $\kappa = 1$ if that denominator degenerates; no other
hyperparameter is tuned. The prior mean is the MIPS percentile rank within $C$:
because the pool is returned MIPS-sorted, this is simply
$\mu_0 = (N - 1 - \mathrm{pos})/(N - 1)$ for a document at position $\mathrm{pos}$.
The query enters as an extra row of the kernel, which is therefore the augmented
$(N+1) \times (N+1)$ matrix with query entries $\exp(\kappa(\cos(q, d) - 1))$, and
is conditioned on at $y = 1$ before the first CE call is spent.

Each posterior update takes the Cholesky factor of
$\widetilde{K}_t = K(A_t, A_t) + \sigma_n^2 I$, with jitter $\sigma_n^2 = 10^{-4}$
added to the observed-block diagonal, and performs two triangular solves: one for
$\alpha = \widetilde{K}_t^{-1}(y_t - \mu_0(A_t))$ and one for
$v = \widetilde{K}_t^{-1}\mathbf{k}(A_t, U)$ over the unobserved set $U$. The
posterior is then $\mu_{\mathrm{post}} = \mu_0 + \mathbf{k}^\top \alpha$ and
$\sigma^2 = \mathrm{diag}(K)_U - \sum_i \mathbf{k}_{iU} v_{iU}$, with the variance
clipped at $0$. If the factorization fails, the jitter is escalated to $10^{-2}$
and, failing that, the solve falls back to a pseudo-inverse. As described in
\S\ref{sec:prior}, the quantity conditioned on is the residual against $\mu_0$,
Z-standardized before conditioning and de-standardized afterwards
(Algorithm~\ref{alg:posterior}). When ranking the pool by $\mu$, documents with
equal posterior means are ordered by MIPS percentile descending.

The resulting costs are $O(N^2)$ for the one-time kernel build, $O(t^2 N)$ per
posterior update and $O(B^3 N)$ over the run, and $O(k(N-k))$ per acquisition step
and $O(B k (N-k))$ over the run; Appendix~\ref{sec:app-complexity} derives these.

\subsection{Kernel ablation}
\label{sec:app-kernel}

To ensure robustness of results with respect to the kernel used in the GP, we run
both GP methods (BoundaryMORPH and BAGEL) with three different kernels. We test
the vMF kernel, described in the Methods section, along with the
squared-exponential (RBF) and Mat\'ern-5/2 kernels
\citep{williams_gaussian_1995}. Both alternatives are functions of the squared
chordal distance $\lVert d - d' \rVert^2 = 2\,(1 - \cos(d,d'))$ on the unit
sphere, a monotone reparametrization of the same cosine geometry the vMF kernel
uses. Writing $u = 1 - \cos(d,d')$ and $\rho = \sqrt{5u/\ell^2}$,
\begin{equation}
K_{\mathrm{RBF}}(d,d') = \exp\!\big(-u/\ell^{2}\big), \qquad
K_{\mathrm{Mat}}(d,d') = \Big(1 + \rho + \tfrac{\rho^{2}}{3}\Big)\,e^{-\rho}.
\label{eq:alt-kernels}
\end{equation}

The three differ only in how fast covariance decays with angle, and thus in the
smoothness they assume of the CE surface: vMF and RBF are infinitely
differentiable, whereas Mat\'ern-5/2 is only twice differentiable and so admits a
rougher residual field. The length-scale is set by the same median heuristic used
for $\kappa$ in the vMF case,
$\ell^2 = \mathrm{median}_{i<j}(1 - \cos(d_i,d_j))$, so no kernel receives a
tuning advantage. Jitter, the query-as-observation anchor (which enters through
the same kernel), residual standardization, and the pool are unchanged.

We fix acquisition and prior for each of the two methods --- so the only thing
that varies is the kernel --- and show the change in performance from switching to
RBF or Mat\'ern-5/2 over the vMF kernel in Table~\ref{tab:kernel-ablation}. Both
metrics are reported, since the two answer different questions
(\S\ref{sec:metrics}) and a kernel could in principle help one and hurt the other.
Each query's score is averaged over the two budgets $B \in \{25, 50\}$ before the
vMF score is subtracted, and the paired tests are run on those per-query
$B$-averaged values. BAGEL's $\beta_0$ is re-selected within each kernel, so
neither alternative is handicapped by a choice tuned for vMF.

The table shows that choice of kernel has no significant effect. Every one of the
72 deltas is under 1.5 pp and the median is 0.32 pp, and both the signs and the
ordering of the two alternatives are inconsistent across datasets: Mat\'ern-5/2 is
nominally best for BoundaryMORPH on NeuCLIR at CE-nCG@50 ($+0.43$ pp) but RBF
leads there on Robust04 ($+0.44$ vs.\ $+0.17$ pp), and on TravelDest both drift
slightly negative at the larger cutoffs. Across the 72 paired Wilcoxon
signed-rank tests behind the table, 13 reach $p < 0.05$ --- against $\approx 4$
expected by chance --- and after Benjamini-Hochberg correction \textbf{5} survive
at $q < 0.05$. All five are BAGEL, none our method: four are BAGEL under RBF
performing \emph{worse} than under vMF (NeuCLIR, $-1.33$ pp CE-nCG@200 and $-1.48$
pp nCG@200; Robust04, $-0.58$ pp CE-nCG@100 and $-0.74$ pp CE-nCG@200), and the
fifth is Mat\'ern-5/2 \emph{helping} BAGEL on TravelDest ($+0.67$ pp CE-nCG@200).
So the only detectable kernel effects move the baseline, in both directions, and
none is large enough to matter: for scale, the same-grid gap between the two
\emph{methods} at a fixed vMF kernel is $+4.8$ to $+10.4$ pp CE-nCG. The
\emph{smallest} method effect is more than three times the largest kernel effect
anywhere in the table, and fifteen times the median one. Selection skill in this
problem is governed by the acquisition rule and the prior, not by the covariance
shape. Hence, we report the paper's main results using the vMF kernel.

\begin{table}
\centering
\scriptsize
\caption{Kernel ablation: $\Delta$ CE-nCG from replacing the vMF kernel (kernel $-$
vMF, percentage points), by dataset and evaluation cutoff $k$. Positive = the
alternative kernel is better. $\dagger$ marks the 5/72 cells significant after
multiple-comparison correction (see text).}
\label{tab:kernel-ablation}
\begin{tabular}{ll rrr !{\hskip 0.4em\vrule\hskip 0.4em} rrr}
\toprule
 & & \multicolumn{3}{c}{$\Delta$ CE-nCG@$k$ (pp)} & \multicolumn{3}{c}{$\Delta$ nCG@$k$ (pp, human qrels)} \\
\cmidrule(lr){3-5}\cmidrule(lr){6-8}
Method & Kernel & 50 & 100 & 200 & 50 & 100 & 200 \\
\midrule
\multicolumn{8}{l}{\emph{NeuCLIR}} \\
\quad \textbf{MORPH} & RBF & $+0.12$ & $+0.29$ & $-0.15$ & $+0.01$ & $+0.22$ & $+0.03$ \\
\quad \textbf{MORPH} & Mat\'ern-5/2 & $+0.43$ & $+0.08$ & $+0.15$ & $+1.23$ & $-0.36$ & $+0.20$ \\
\quad BAGEL & RBF & $-0.61$ & $-0.71$ & $-1.33^{\dagger}$ & $-0.60$ & $-0.99$ & $-1.48^{\dagger}$ \\
\quad BAGEL & Mat\'ern-5/2 & $-0.12$ & $+0.28$ & $+0.31$ & $-0.55$ & $-0.79$ & $-0.79$ \\
\midrule
\multicolumn{8}{l}{\emph{Robust04}} \\
\quad \textbf{MORPH} & RBF & $+0.44$ & $+0.19$ & $+0.30$ & $+0.76$ & $+0.55$ & $+0.08$ \\
\quad \textbf{MORPH} & Mat\'ern-5/2 & $+0.17$ & $+0.38$ & $+0.33$ & $+0.44$ & $+0.81$ & $+0.11$ \\
\quad BAGEL & RBF & $+0.01$ & $-0.58^{\dagger}$ & $-0.74^{\dagger}$ & $+0.13$ & $-0.65$ & $-0.67$ \\
\quad BAGEL & Mat\'ern-5/2 & $+0.44$ & $+0.30$ & $+0.23$ & $+0.57$ & $+0.44$ & $+0.32$ \\
\midrule
\multicolumn{8}{l}{\emph{TravelDest}} \\
\quad \textbf{MORPH} & RBF & $+0.06$ & $-0.04$ & $-0.42$ & $+0.81$ & $-0.33$ & $+0.14$ \\
\quad \textbf{MORPH} & Mat\'ern-5/2 & $+0.10$ & $-0.13$ & $-0.09$ & $-0.22$ & $+0.17$ & $-0.25$ \\
\quad BAGEL & RBF & $+0.12$ & $-0.06$ & $-0.13$ & $+0.22$ & $-0.32$ & $-0.26$ \\
\quad BAGEL & Mat\'ern-5/2 & $+0.16$ & $+0.39$ & $+0.67^{\dagger}$ & $+0.46$ & $+0.37$ & $+0.47$ \\
\bottomrule
\end{tabular}
\end{table}

\section{Experimental Setup}
\label{sec:app-setup}

\subsection{Dataset statistics}
\label{sec:app-datasets}

\textbf{NeuCLIRBench} \citep{lawrie_neuclirbench_2025} pairs open-ended English
information needs, each stated as a keyword line followed by a question, against a
newswire corpus machine-translated into English from Persian, Russian and Chinese.

\textbf{Robust04} \citep{voorhees_trec_2005} uses TREC topics, each a
description-style sentence, over the TREC Disks 4 \& 5 newswire and government
collection.

\textbf{TravelDest} \citep{wen_elaborative_2024} uses broad travel-intent topics
naming a desired quality of a destination rather than a place, over
Wikivoyage-derived passages describing individual cities.

\begin{table}
\centering
\small
\caption{Active-testbed statistics: query counts, relevant documents per query
(showing these queries are genuinely diffuse), MIPS pool coverage, and document
length, per dataset. Definitions and the query filter are given below.}
\label{tab:dataset-stats}
\begin{tabular}{l rrr}
\toprule
& NeuCLIR & Robust04 & TravelDest \\
\midrule
Full-corpus docs & 10,038,768 & 528,155 & 131,268 \\
Substrate pool / query & 10,000 & 10,000 & 10,000 \\
Substrate union docs & 731,335 & 422,904 & 90,727 \\
\# Queries (original) & 105 & 249 & 100 \\
\# Queries (post-filter) & 85 & 187 & 99 \\
\# docs/query: median & 51 & 62 & 462 \\
\# docs/query: mean & 70.3 & 89.2 & 839.5 \\
\# docs/query: IQR (p25--p75) & 37--85 & 35--116 & 140--1,062 \\
\# docs/query: max & 246 & 448 & 5,667 \\
Pool coverage: mean & 0.92 & 0.88 & 0.70 \\
Pool coverage: median & 0.96 & 0.92 & 0.74 \\
Pool coverage: p10 & 0.81 & 0.73 & 0.47 \\
Pool coverage: min & 0.58 & 0.42 & 0.29 \\
Doc tokens: median & 335 & 426 & 68 \\
Doc tokens: IQR (p25--p75) & 218--517 & 226--805 & 38--109 \\
\bottomrule
\end{tabular}
\end{table}

Relevance basis: NeuCLIR qrels grade $> 0$ (grades in $\{0,1,3\}$); Robust04 qrels
grade in $\{1,2\}$; TravelDest an explicit binary relevant set. Pool coverage =
fraction of a query's judged-relevant documents present in its top-10,000 MIPS
pool, macro-averaged, denominator = all judged-relevant documents in the full
qrels, so $1 -$ coverage is relevant mass discarded before any selection and
bounds what any selector could recover. Token counts are Qwen3-Embedding-4B tokens
over the pool union. The $\ge 20$ filter drops 20 / 62 / 1 queries.

\begin{table}
\centering
\small
\caption{Relevant documents per retained query, full percentiles.}
\label{tab:app-rel-percentiles}
\begin{tabular}{l rrrrrrrr}
\toprule
Dataset & min & p10 & p25 & median & mean & p75 & p90 & max \\
\midrule
NeuCLIR & 22 & 27 & 37 & 51 & 70.3 & 85 & 122 & 246 \\
Robust04 & 20 & 27 & 35 & 62 & 89.2 & 116 & 194 & 448 \\
TravelDest & 20 & 67 & 140 & 462 & 839.5 & 1,062 & 1,975 & 5,667 \\
\bottomrule
\end{tabular}
\end{table}

\begin{table}
\centering
\small
\caption{Pool coverage, full percentiles. All columns are macro-averaged over queries except the last, which is the micro-averaged coverage (pooled over all judged-relevant documents).}
\label{tab:app-coverage-percentiles}
\begin{tabular}{l rrrrrrrr r}
\toprule
Dataset & min & p10 & p25 & median & mean & p75 & p90 & max & micro \\
\midrule
NeuCLIR & 0.582 & 0.810 & 0.897 & 0.957 & 0.922 & 0.981 & 1.000 & 1.000 & 0.904 \\
Robust04 & 0.422 & 0.729 & 0.831 & 0.917 & 0.885 & 0.975 & 1.000 & 1.000 & 0.860 \\
TravelDest & 0.287 & 0.466 & 0.607 & 0.737 & 0.699 & 0.821 & 0.882 & 0.943 & 0.599 \\
\bottomrule
\end{tabular}
\end{table}

Relevant documents in the qrels against relevant documents inside the pool:
5,976 $\to$ 5,400 (NeuCLIR), 16,684 $\to$ 14,351 (Robust04),
83,112 $\to$ 49,822 (TravelDest).

\begin{table}
\centering
\small
\caption{Document tokens over the pool union, full percentiles.}
\label{tab:app-token-percentiles}
\begin{tabular}{l rrrrrrr}
\toprule
Dataset & min & p25 & median & mean & p90 & p99 & max \\
\midrule
NeuCLIR & 3 & 218 & 335 & 444.0 & 824 & 2,261 & 18,053 \\
Robust04 & 15 & 226 & 426 & 639.1 & 1,264 & 3,299 & 927,351 \\
TravelDest & 6 & 38 & 68 & 81.4 & 157 & 289 & 918 \\
\bottomrule
\end{tabular}
\end{table}

\subsection{Detailed baselines}
\label{sec:app-baselines}

All baselines are scored on the \textbf{same fixed per-query top-10,000 pool} as
BoundaryMORPH, so every comparison isolates allocation and ordering within a
shared candidate set rather than retrieval from scratch. Every baseline that has
hyperparameters has them \textbf{selected on the evaluation metric} (pooled-mean
CE-nCG@50) per cell per budget, and reused across $k$; BoundaryMORPH runs at one
fixed configuration and tunes nothing.

\textbf{BM25.} Okapi BM25 with $k_1 = 1.5$, $b = 0.75$ and full-corpus
statistics, applied as a re-ranking of the shared pool and truncated at $k$. It
makes no CE calls, so its ordering is fixed per query and its value is identical at
every $B$; and its ordering uses neither MIPS nor the CE, so its
nCG is a single value per (dataset, $k$) and its CE-nCG depends only on which CE
supplies the gold. Both properties are verified rather than assumed.

\textbf{SS (single-stage rerank).} Score the MIPS top-$B$ with the CE, place
those $B$ above everything else, and leave the tail in MIPS order. For $k \ge B$
this returns exactly the MIPS top-$k$, so the budget buys nothing in that regime.

\textbf{BAGEL.} Zero-prior GP-UCB. It shares our pool, our vMF kernel, and our
median-heuristic bandwidth, and differs in exactly three places: the prior mean
(zero rather than MIPS percentile), the acquisition rule (UCB over the whole pool
rather than the boundary), and the warm-start.

\begin{itemize}
\item $\beta_0$ swept over $\{0, 0.5, 1, 2\}$ with square-root annealing
  $\beta_t = \beta_0/\sqrt{t+1}$, selected per (dataset $\times$ embedder $\times$
  CE) cell per $B$. $\beta_0 = 0$ --- pure exploitation of the posterior mean ---
  is selected in 4 of the headline cells, all at large $B$.
\item BAGEL is \textbf{warm-started}: it spends $\lfloor B/2 \rfloor$ calls
  deterministically on the MIPS top-$\lfloor B/2 \rfloor$, leaving
  $B - \lfloor B/2 \rfloor$ acquisition steps.
\end{itemize}

\textbf{RGS.} Greedy graph expansion under a max-heap keyed by CE score: pop
the highest-scored not-yet-expanded document, fetch its graph neighbors, score any
that are unscored, push them, and truncate the heap to $L_s$. The graph is a
DiskANN cosine graph over the pool embeddings (one greedy search + robust prune
pass, $L = 100$, $\alpha = 1.2$).

\begin{itemize}
\item Swept: graph degree, which is also the expansion branching factor,
  $R \in \{10, 32, 64\}$; the seed/expansion split
  $n_{\mathrm{seeds}} = \mathrm{round}(B / \mathrm{seed\_div})$; and the heap cap
  $L_s = \max(n_{\mathrm{seeds}}, \mathrm{round}(\mathrm{frac} \cdot B))$ with
  $\mathrm{frac} \in \{0.2, 0.5, 1.0\}$. All headline cells select $R = 10$,
  $\mathrm{seed\_div} = 2$, $\mathrm{frac} = 1.0$. The $\mathrm{seed\_div}$ grid is
  $\{5, 4, 3, 2\}$ on the main configuration and $\{2, 4, 8\}$ on the model-swap
  configurations; the selected value, 2, is in both.
\item Ranking convention: scored documents receive
  $\mathrm{offset} + \text{CE score}$ with $\mathrm{offset} = 10$ and unscored
  receive 0, so scored documents form a band strictly above unscored and a stable
  descending sort leaves the unscored tail in MIPS order. This is the most
  charitable construction of the tail for a method that cannot rank what it never
  scored.
\item At $n_{\mathrm{seeds}} = B$ the search reduces to single-stage rerank to
  within $10^{-12}$, which checks the implementation against its own degenerate
  case.
\end{itemize}

\textbf{Agentic loop.} A multi-hop search agent (DSP) run with plain
instruction prompts, generated by Qwen3.5-9B with thinking disabled and greedy
decoding. Retrieval is \emph{within-pool}, so the loop is scored on the same
candidate set as the selectors. Each hop the model writes one query from the
information need and its accumulated notes, that query is embedded with the same
retriever and its top documents by within-pool cosine are shown, and the model
scores each document 0--10 and records new notes. Documents shown in earlier hops
are excluded from later ones (meaning the agent always scores \emph{exactly}
docs/hop new documents in each hop, and \emph{exactly}
docs/hop $\times$ hops total documents). The final ranking is surfaced documents by
assigned score descending, ties broken by cosine to the original query, followed by
the never-surfaced tail in MIPS order. Grid: per-hop depth
$\in \{5, 10, 20, 50, 100\}$ $\times$ hops $\in \{1, 2, 3, 4, 6, 8\}$.

The query-generation prompt is shown in Figure~\ref{fig:prompt-query-gen}, and the
note-extraction prompt in Figure~\ref{fig:prompt-note-extraction}.

\begin{figure}[h!]
    \centering
    \begin{tcolorbox}[colback=gray!30, colframe=black, boxrule=0.5mm, width=\columnwidth]
    \tiny
    You are a search agent gathering documents to satisfy an information need. Given
    the information need and the notes gathered so far, write ONE new search query
    that will retrieve MORE relevant documents for the information need. This can mean
    exploring a different facet OR retrieving more documents from facets already
    covered --- do whatever will surface the most additional relevant documents. The
    query MUST be at least 100 words long --- write a long, detailed, keyword-rich
    query (at least 100 words), not a short phrase. Return ONLY the query text, no
    preamble, no quotes.

    Claim: \texttt{\{claim\}}

    Notes so far:
    \texttt{\{notes\}}

    New search query:
    \end{tcolorbox}
    \caption{Prompt for generating the next search query in the agentic retrieval loop (DSP baseline).}
    \label{fig:prompt-query-gen}
\end{figure}

\begin{figure}[h!]
    \centering
    \begin{tcolorbox}[colback=gray!30, colframe=black, boxrule=0.5mm, width=\columnwidth]
    \tiny
    You are a search agent. Below is the information need, the notes gathered so far,
    and a set of newly retrieved documents (each labelled
    \texttt{[DOC <index>]}). For EACH document, assign a relevance SCORE from 0 to 10
    (0 = irrelevant to the information need, 10 = perfectly relevant). Then extract
    any NEW facts from the relevant documents that help satisfy the information need.

    Respond with STRICT JSON ONLY (no markdown fence, no prose) of the form:
    \texttt{\{"doc\_scores": \{"0": 8, "3": 2\}, "new\_notes": ["...", "..."]\}}
    Put doc\_scores FIRST. Use the EXACT integer indices shown in the
    \texttt{[DOC <index>]} labels as the keys, and an INTEGER 0-10 as each value.
    Score EVERY document shown. Keep each note to ONE short sentence.

    Claim: \texttt{\{claim\}}

    Notes so far:
    \texttt{\{notes\}}

    Retrieved documents:
    \texttt{\{context\}}

    JSON:
    \end{tcolorbox}
    \caption{Prompt for scoring retrieved documents and extracting notes in the agentic retrieval loop (DSP baseline).}
    \label{fig:prompt-note-extraction}
\end{figure}

Across all 29 configurations, NeuCLIR, $n = 85$, nCG@100 ranges from 0.554 (5
documents per hop, 1 hop, 57.6 s) to 0.618 (100 per hop, 3 hops, 884.6 s), and
latency from 57.6 s to 1,922.8 s. BoundaryMORPH reaches nCG@100 0.621 in 25.7 s at
$B = 100$, $N = 10{,}000$, so no configuration of the loop matches it on either
quality or cost.

\section{Additional Results and Robustness}
\label{sec:app-results}

\subsection{Full ablation results}
\label{sec:app-ablation-full}

Table~\ref{tab:ablation-full} breaks the ablation of \S\ref{sec:ablations-analysis} out by dataset and reports both nCG@$k$ and CE-nCG@$k$, with values averaged over the four embedder$\times$CE pairs. Removing either component lowers every cell.

\begin{table}
\centering
\scriptsize
\caption{Full BoundaryMORPH ablations, per dataset and for both metrics (Table~\ref{tab:ablation} in the main text averages the nCG@$k$ columns over datasets). Each row independently replaces one component: boundary acquisition$\to$ standard UCB; MIPS prior$\to$ 0 prior. We report the pp change from full BoundaryMORPH. Both components are crucial for the gains, and \textit{always} improve results.}
\label{tab:ablation-full}
\begin{tabular}{l rrr rrr !{\hskip 0.4em\vrule\hskip 0.4em} rrr rrr}
\toprule
 & \multicolumn{6}{c}{nCG@$k$ (\%, human qrels)} & \multicolumn{6}{c}{CE-nCG@$k$ (\%)} \\
\cmidrule(lr){2-7}\cmidrule(lr){8-13}
 & \multicolumn{3}{c}{$B$=25} & \multicolumn{3}{c}{$B$=50} & \multicolumn{3}{c}{$B$=25} & \multicolumn{3}{c}{$B$=50} \\
\cmidrule(lr){2-4}\cmidrule(lr){5-7}\cmidrule(lr){8-10}\cmidrule(lr){11-13}
\diagbox[width=8em,height=2.2em]{Method}{$k$} & 50 & 100 & 200 & 50 & 100 & 200 & 50 & 100 & 200 & 50 & 100 & 200 \\
\midrule
\multicolumn{1}{c}{\emph{NeuCLIR}} \\
\textbf{MORPH} & 51.2 & 56.4 & 64.4 & 54.8 & 59.7 & 66.6 & 84.6 & 80.8 & 78.8 & 89.9 & 85.3 & 82.4 \\
\quad w/o acq. & \cellcolor[rgb]{1.000,0.792,0.792} -2.1 & \cellcolor[rgb]{1.000,0.791,0.791} -2.1 & \cellcolor[rgb]{1.000,0.872,0.872} -1.3 & \cellcolor[rgb]{1.000,0.876,0.876} -1.2 & \cellcolor[rgb]{1.000,0.730,0.730} -2.7 & \cellcolor[rgb]{1.000,0.844,0.844} -1.5 & \cellcolor[rgb]{1.000,0.604,0.604} -3.9 & \cellcolor[rgb]{1.000,0.478,0.478} -5.2 & \cellcolor[rgb]{1.000,0.536,0.536} -4.6 & \cellcolor[rgb]{1.000,0.739,0.739} -2.6 & \cellcolor[rgb]{1.000,0.481,0.481} -5.2 & \cellcolor[rgb]{1.000,0.486,0.486} -5.1 \\
\quad w/o prior & \cellcolor[rgb]{1.000,0.765,0.765} -2.3 & \cellcolor[rgb]{1.000,0.565,0.565} -4.3 & \cellcolor[rgb]{1.000,0.450,0.450} -5.5 & \cellcolor[rgb]{1.000,0.847,0.847} -1.5 & \cellcolor[rgb]{1.000,0.700,0.700} -3.0 & \cellcolor[rgb]{1.000,0.707,0.707} -2.9 & \cellcolor[rgb]{1.000,0.705,0.705} -2.9 & \cellcolor[rgb]{1.000,0.604,0.604} -3.9 & \cellcolor[rgb]{1.000,0.561,0.561} -4.4 & \cellcolor[rgb]{1.000,0.773,0.773} -2.3 & \cellcolor[rgb]{1.000,0.764,0.764} -2.3 & \cellcolor[rgb]{1.000,0.724,0.724} -2.7 \\
\midrule
\multicolumn{1}{c}{\emph{Robust04}} \\
\textbf{MORPH} & 46.4 & 48.2 & 55.1 & 51.4 & 52.0 & 57.7 & 69.8 & 67.9 & 68.1 & 76.1 & 72.3 & 71.1 \\
\quad w/o acq. & \cellcolor[rgb]{1.000,0.659,0.659} -3.4 & \cellcolor[rgb]{1.000,0.653,0.653} -3.5 & \cellcolor[rgb]{1.000,0.727,0.727} -2.7 & \cellcolor[rgb]{1.000,0.794,0.794} -2.0 & \cellcolor[rgb]{1.000,0.747,0.747} -2.5 & \cellcolor[rgb]{1.000,0.845,0.845} -1.5 & \cellcolor[rgb]{1.000,0.656,0.656} -3.4 & \cellcolor[rgb]{1.000,0.638,0.638} -3.6 & \cellcolor[rgb]{1.000,0.712,0.712} -2.9 & \cellcolor[rgb]{1.000,0.774,0.774} -2.2 & \cellcolor[rgb]{1.000,0.672,0.672} -3.3 & \cellcolor[rgb]{1.000,0.725,0.725} -2.7 \\
\quad w/o prior & \cellcolor[rgb]{1.000,0.647,0.647} -3.5 & \cellcolor[rgb]{1.000,0.676,0.676} -3.2 & \cellcolor[rgb]{1.000,0.583,0.583} -4.1 & \cellcolor[rgb]{1.000,0.758,0.758} -2.4 & \cellcolor[rgb]{1.000,0.778,0.778} -2.2 & \cellcolor[rgb]{1.000,0.723,0.723} -2.8 & \cellcolor[rgb]{1.000,0.587,0.587} -4.1 & \cellcolor[rgb]{1.000,0.594,0.594} -4.0 & \cellcolor[rgb]{1.000,0.583,0.583} -4.1 & \cellcolor[rgb]{1.000,0.681,0.681} -3.2 & \cellcolor[rgb]{1.000,0.683,0.683} -3.1 & \cellcolor[rgb]{1.000,0.709,0.709} -2.9 \\
\midrule
\multicolumn{1}{c}{\emph{TravelDest}} \\
\textbf{MORPH} & 43.7 & 39.2 & 38.2 & 47.8 & 42.1 & 40.3 & 68.2 & 67.5 & 68.7 & 75.1 & 72.1 & 71.8 \\
\quad w/o acq. & \cellcolor[rgb]{1.000,0.774,0.774} -2.2 & \cellcolor[rgb]{1.000,0.816,0.816} -1.8 & \cellcolor[rgb]{1.000,0.778,0.778} -2.2 & \cellcolor[rgb]{1.000,0.874,0.874} -1.3 & \cellcolor[rgb]{1.000,0.835,0.835} -1.6 & \cellcolor[rgb]{1.000,0.844,0.844} -1.5 & \cellcolor[rgb]{1.000,0.695,0.695} -3.0 & \cellcolor[rgb]{1.000,0.632,0.632} -3.7 & \cellcolor[rgb]{1.000,0.604,0.604} -3.9 & \cellcolor[rgb]{1.000,0.781,0.781} -2.2 & \cellcolor[rgb]{1.000,0.668,0.668} -3.3 & \cellcolor[rgb]{1.000,0.658,0.658} -3.4 \\
\quad w/o prior & \cellcolor[rgb]{1.000,0.682,0.682} -3.2 & \cellcolor[rgb]{1.000,0.819,0.819} -1.8 & \cellcolor[rgb]{1.000,0.810,0.810} -1.9 & \cellcolor[rgb]{1.000,0.751,0.751} -2.5 & \cellcolor[rgb]{1.000,0.840,0.840} -1.6 & \cellcolor[rgb]{1.000,0.860,0.860} -1.4 & \cellcolor[rgb]{1.000,0.544,0.544} -4.5 & \cellcolor[rgb]{1.000,0.586,0.586} -4.1 & \cellcolor[rgb]{1.000,0.636,0.636} -3.6 & \cellcolor[rgb]{1.000,0.549,0.549} -4.5 & \cellcolor[rgb]{1.000,0.562,0.562} -4.4 & \cellcolor[rgb]{1.000,0.640,0.640} -3.6 \\
\bottomrule
\end{tabular}
\end{table}

\subsection{Robustness to retriever and cross-encoder}
\label{sec:app-robustness}

\begin{figure}
    \centering
    \includegraphics{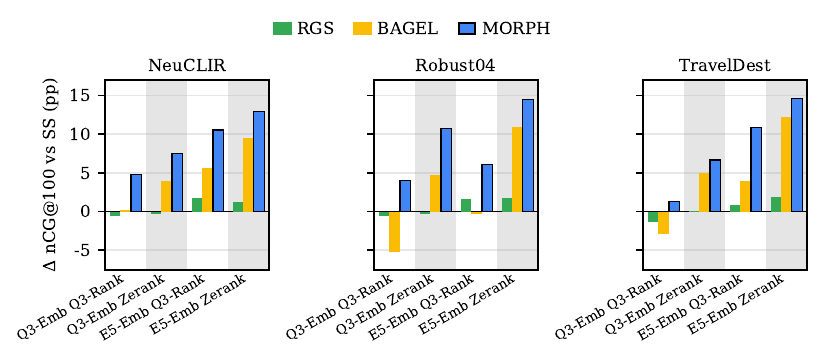}
    \caption{\textbf{The advantage is robust to choice of retriever and
    cross-encoder.} Per-dataset nCG@100 lift over single-stage rerank (percentage points)
    at $B = 50$, for all four embedder $\times$ CE combinations. BoundaryMORPH holds a
    gap over every baseline in all four, on every dataset.}
    \label{fig:embedder-ablation-delta}
\end{figure}

\textbf{Results are robust to the choice of retriever and cross-encoder.}\label{sec:robustness} To test our method's robustness to embedding model and CE model, we use two
different embedding models (Qwen3-Embedding-4B \citep{zhang_qwen3_2025} and
multilingual-e5-large \citep{wang_multilingual_2024}), and two different CE models
(Qwen3-Reranker-8B \citep{zhang_qwen3_2025} and zerank-2
\citep{pipitone_zelo_2025}) and test all methods with all 4 combinations.
Figure~\ref{fig:embedder-ablation-delta} (Appendix~\ref{sec:app-per-dataset}) shows the per-dataset result at $B=50$, $k=100$:
BoundaryMORPH maintains a gap over baselines with all embedder/CE combinations on every dataset,
indicating that the result is not a property of the retriever and CE.

\textbf{Gains are largest with a weak retriever.} Mean lift over SS runs from
\textbf{+3.4 pp} with a strong retriever and strong CE to \textbf{+14.0 pp} with a
weaker retriever; the advantage grows as retriever weakens, likely because there
is simply more headroom between retriever and CE to take advantage of. Though in \textit{absolute} terms, all methods improve when switching from the weaker e5 embedder to the stronger Qwen3 model, averaging $+10.6$ nCG@100 (B=50) for SS, an $+8.6$ for RGS, and $+4.9$ for BoundaryMORPH; a stronger prior universally helps.

\subsection{Per-dataset embedder $\times$ CE results}
\label{sec:app-per-dataset}

nCG@100 lift over single-stage rerank (pp) at $B = 50$, per dataset, for all four
embedder $\times$ CE combinations. Embedders: Qwen3-Emb = Qwen3-Embedding-4B,
e5 = multilingual-e5-large. CEs: Qwen-Rerank = Qwen3-Reranker-8B,
zerank = zerank-2.

\subsection{Latency Analysis}
\label{sec:latency}

\begin{wrapfigure}{R}{2.6in}
    \centering
    \includegraphics[width=2.5in]{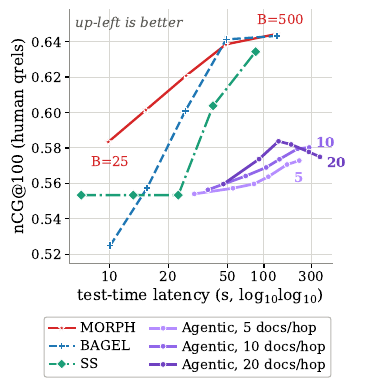}
    \caption{\textbf{Budget is a realistic proxy for latency.} We show nCG@100
    against wall-clock latency per query on NeuCLIR, including an agentic loop (DSP). Latency scales with budget for all methods; with performance converging at high $B$. The agentic method is dominated due to decoding latency.}
    \label{fig:latency}
\end{wrapfigure}

Selectors that all call the same CE can be compared via $B$. We also compare wall
clock latency (Figure~\ref{fig:latency}, run on one A10, on NeuCLIR), and include an agentic retrieval loop in
the comparison. We use a plain multi-hop retrieve-and-read loop, DSP
\citep{khattab_demonstrate-search-predict_2023}\ct{, \citep{khattab_dspy_2024}}. The model (Qwen3.5-9B, running on 8 A10s) writes a search
query, retrieves, reads and scores the results, records notes, and uses its notes
to write another query (and this is repeated). Prompts and other details are in
Appendix~\ref{sec:app-baselines} and Appendix~\ref{sec:app-latency}.

While such results are highly dependent on document length and hardware, figure~\ref{fig:latency} shows that CE budget scales each method's latency,
justifying the use of $B$ as a latency dial. BoundaryMORPH nearly
Pareto-dominates, except at high $B$ ($B>>k$). As $B$ increases, selection matters less;
there is budget enough to score every plausible document, so every reasonable
allocator lands on a similar set. Thus, most methods converge at high $B$.

Due to the GP, BoundaryMORPH and BAGEL must be run sequentially, while SS can be batched. In practice we find that on 1 GPU, the variable document length leads to extra batch padding which \textit{increases} latency for SS at batch sizes $\{2, 3, 8, 16,32,64\}$, so we report results on batch size of 1, which is the fastest possible setting for SS. In theory, with enough GPUs/memory, SS can be fully amortized, though based on the empirical results for our datasets, real-time speed would improve only from added GPUs, and not from batching on any given GPU. Thus, at least in theory, the latency of SS could be improved in an exact linear relationship with the number of GPUs.

Notably, despite $8 \times$ parallel compute, the agentic loop is Pareto-dominated. The whole agentic family lies below
the budgeted frontier at every latency it reaches. Latency grows close to linearly
in hops, because each hop pays for autoregressive decoding.

\textbf{Origin of the gap.} The information the GP exploits is already latent
in the corpus embedding geometry and propagates through the kernel to documents
never read. This agentic loop can only carry forward what it wrote down in natural
language, in a bounded ``notes'', and pays decoding latency per token. We
hypothesize that the gap is explained by the GP being much more efficient at
propagating relevance information.

\subsection{Latency setup}
\label{sec:app-latency}

\textbf{Hardware.} One A10G (24 GB), bf16 with SDPA, left padding, maximum
length 4,096. Document sequence lengths are p50 = 426, p95 = 1,217 tokens.
Absolute seconds are hardware-specific; the ratios are the claim.

\textbf{Per-component breakdown.} NeuCLIR, $N = 10{,}000$, capacity 50,
$n = 85$, medians (Table~\ref{tab:app-latency}).

\begin{table}
\centering
\small
\caption{Median per-query latency by component. NeuCLIR, $N = 10{,}000$, capacity 50, $n = 85$. \emph{algorithm} is the total non-CE time; \emph{GP update} and \emph{acquisition} are its components.}
\label{tab:app-latency}
\begin{tabular}{r rr rrr}
\toprule
$B$ & CE & CE \% & algorithm & GP update & acquisition \\
\midrule
25 & 4.66 s & 95\% & 0.22 s & 0.06 s & 0.16 s \\
50 & 9.43 s & 95\% & 0.51 s & 0.17 s & 0.35 s \\
100 & 19.45 s & 93\% & 1.46 s & 0.63 s & 0.85 s \\
200 & 38.89 s & 90\% & 4.56 s & 2.73 s & 1.78 s \\
500 & 94.39 s & 74\% & 32.86 s & 27.86 s & 4.79 s \\
\midrule
\multicolumn{6}{l}{\emph{Totals:} 4.88 s / 9.94 s / 20.91 s / 43.45 s / 127.25 s at $B = 25/50/100/200/500$.} \\
\bottomrule
\end{tabular}
\end{table}

\begin{itemize}
\item CE dominates latency at every $B$, 74--95\%; the share is highest at
  small $B$ (95\% at $B = 25, 50$) and falls only at $B = 500$ (74\%), where
  GP update cost grows enough to compete.
\item The algorithm's own cost (GP update + acquisition) is under 1.5 s for
  $B \le 100$ and grows to 32.86 s at $B = 500$, driven almost entirely by the
  GP update.
\item Per-call CE cost is ${\sim}0.19$ s, flat in $B$.
\end{itemize}

\section{Query Multimodality}
\label{sec:app-multimodality}

Intuitively, diffuse queries can be \emph{unimodal}, where relevant documents are
relatively similar to one another, or \emph{multimodal}, where relevant documents
are relatively different from one another \citep{chen_beyond_2025,
song_multi3ir_2026, liu_multimodal_2026}. We expect multimodal queries to be
especially difficult for max-seeking selectors, which concentrate calls on the
primary mode, and define a geometric scalar, \textit{multimodal mass}, to measure
this.

\subsection{Peak extraction and LLM audit}
\label{sec:app-peaks}

\begin{table}
\centering
\small
\caption{Adjusted Rand index between the LLM's two-group partition and the geometric peak-of-origin labels, by slice.}
\label{tab:app-ari}
\begin{tabular}{l rrr}
\toprule
Slice & $n$ & mean ARI & median ARI \\
\midrule
All pairs & 496 & 0.635 & 1.000 \\
Pairs including the main peak & 190 & 0.486 & 0.437 \\
Secondary--secondary pairs & 306 & 0.726 & 1.000 \\
Per-query mean & 65 & 0.593 & 0.624 \\
\bottomrule
\end{tabular}
\end{table}

\textbf{Peak extraction.} We adapt the persistence-based clustering of
\citet{chazal_persistence-based_2013}, viewing each document's CE score as its
\emph{height} over the embedding kNN landscape. A \emph{peak} is a group of close,
highly relevant documents, and its \emph{prominence} measures how independent it is
from other peaks: how far relevance must drop before it merges into a taller one.
Concretely, on the top-$M = 1{,}000$ documents by CE score within
each query's $N = 10{,}000$ pool, build a $k_{\mathrm{nn}} = 10$ OR-symmetrized
cosine kNN graph. Sweep a descending CE-score threshold; each connected component
of the super-level set carries a peak = its highest-scoring document, and merging
components record 0-dimensional persistence (merge-tree prominence). Simplifying
at $\tau = 0.20$ yields a hard partition of the top-$M$ into peaks. Multimodal
mass is the sum of mass-weighted prominence over all non-main peaks, so a unimodal
landscape scores 0.

\begin{figure}[b]
    \centering
    \begin{tcolorbox}[colback=gray!30, colframe=black, boxrule=0.5mm, width=\columnwidth]
    \tiny
    You are analyzing search results for a query. Below are \texttt{\{n\}} documents
    that are ALL relevant to the query. They are relevant in different ways --- they
    emphasize different aspects, sub-topics, entities, or angles of the query.

    QUERY: \texttt{\{query\}}

    Your task: partition ALL \texttt{\{n\}} documents into exactly TWO groups based on
    *how* each document is relevant to the query (what aspect/angle/sub-topic of the
    query it addresses). The two groups need NOT be equal in size, but every document
    must be placed in exactly one group, and neither group may be empty. Base the
    split on the substantive way each document relates to the query, not on surface
    features like length or style.

    Give each group a SHORT description (one sentence) of the way its documents are
    relevant to the query.

    DOCUMENTS:
    \texttt{\{docs\}}

    Respond with ONLY a JSON object, no other text, in exactly this form:
    \texttt{\{"group\_1": \{"description": "<one sentence>", "docs": [<doc numbers>]\}, "group\_2": \{"description": "<one sentence>", "docs": [<doc numbers>]\}\}}
    \end{tcolorbox}
    \caption{Prompt for the LLM judge that partitions documents by relevance aspect (peak audit).}
    \label{fig:prompt-peak-judge}
\end{figure}

\textbf{Peaks correspond to semantic distinctions.} To test if peaks are
semantically meaningful, we give an LLM documents drawn from the same peak and
from different peaks, and ask it to partition them, scoring the result against the geometric partition by ARI: \textbf{mean 0.635, median 1.000} over \textbf{496
pairs from 65 multi-peak queries}, \textbf{85\%} above chance and \textbf{63\%}
clean at ARI $\ge 0.5$. The high success indicates that geometric multimodality is
also a semantic feature.

\textbf{LLM audit.} For each query, keep peaks with $\ge 3$ member documents;
for every unordered pair, sample up to 8 documents from each peak's 0.5-core
(falling back to the full basin when the core has $< 3$), shuffle the $\le 16$
together, and ask the judge to split them into two groups by \emph{how} they are
relevant. Score against the peak-of-origin labels by adjusted Rand index. One
sample per pair, seeded deterministically. Judge: Claude Opus 4.5, temperature 0,
max 1,200 tokens, strict-JSON two-group response; documents truncated to 1,500
characters. 496 pairs from 65 multi-peak NeuCLIR queries, parse success 496/496.

ARI histogram: 32 below $-0.05$, 42 at chance $[-0.05, 0.05]$, 46 in
$(0.05, 0.2]$, 40 in $(0.2, 0.4]$, 47 in $(0.4, 0.6]$, 40 in $(0.6, 0.8]$, 249 in
$(0.8, 1]$. 85.1\% beat chance, 62.7\% reach ARI $\ge 0.5$, and half are separated
perfectly. Correlation of ARI with pair size is $-0.18$.

The judge prompt is shown in Figure~\ref{fig:prompt-peak-judge}.

\end{document}

%% file: math_commands.tex
\usepackage{amsmath,amsfonts,bm}

\def\eqref#1{equation~\ref{#1}}

\def\1{\bm{1}}

\DeclareMathAlphabet{\mathsfit}{\encodingdefault}{\sfdefault}{m}{sl}
\SetMathAlphabet{\mathsfit}{bold}{\encodingdefault}{\sfdefault}{bx}{n}

\DeclareMathOperator*{\argmax}{arg\,max}
\DeclareMathOperator*{\argmin}{arg\,min}